# Accelerating Hartree-Fock and Density Functional Theory Calculations using Tensor Hypercontraction

Andreas Erbs Hillers-Bendtsen[1,2], Todd J. Martínez[1,2,*]

[1]Department of Chemistry and The PULSE Institute, Stanford University, Stanford, CA 94305, USA

[2]SLAC National Accelerator Laboratory, 2575 Sand Hill Road, Menlo Park, CA 94025, USA

Email: Todd.Martinez@stanford.edu

**Abstract:** With the widespread use of self-consistent field methods, including Hartree-Fock and Density Functional Theory, the implications of accelerating these methods are immense. To this end, we develop a tensor hypercontraction construction with $O(N^3)$ formal scaling that can accelerate self-consistent field calculations. Using tensor hypercontraction, we implement an empirically $O(N^2)$ scaling Fock matrix construction that is 2-4× faster than existing integral-direct methods, as it avoids the repeated recalculation of two-electron repulsion integrals. In combination with a density-difference ansatz, our tensor hypercontraction self-consistent field implementation tests show errors below $7.0 \times 10^{-3}$ $E_h$ for relative energies on protein systems containing up to 3000 basis functions.

**TOC GRAPHICS**

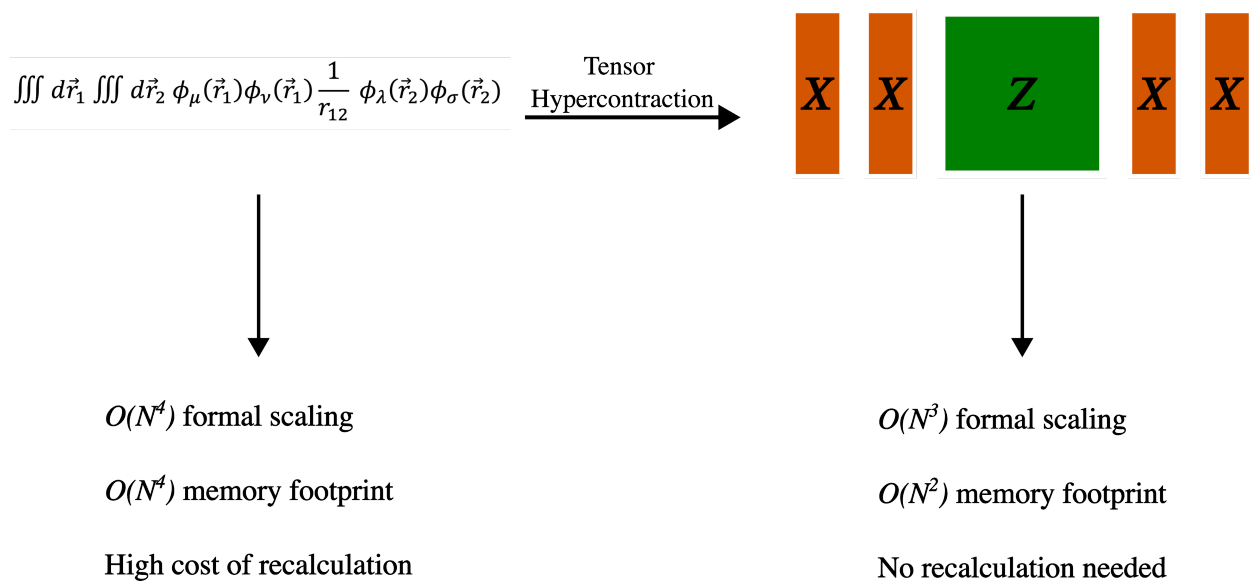

O(N⁴) formal scaling          O(N³) formal scaling

O(N⁴) memory footprint      O(N²) memory footprint

High cost of recalculation     No recalculation needed

**Introduction**

Self-consistent field (SCF) methods such as Hartree-Fock theory[1, 2] continue to play a central role in computational chemistry even though the Hartree-Fock model was introduced almost 100 years ago. Not only does Hartree-Fock theory capture the fundamental physical interactions that lead to chemical bonding and reactivity, but it also provides the theoretical foundation for more accurate treatments of electronic structure such as Møller-Plesset perturbation theory[3] and coupled cluster theory.[4] Perhaps more importantly, the theoretical and algorithmic machinery of Hartree-Fock theory provides the framework for Kohn-Sham density functional theory[5] (DFT). DFT is currently the main workhorse in applied quantum chemistry, since it is formally exact within a one-body picture when the elusive "exact" functional is used. For these reasons, significant effort has been devoted to developing and refining SCF algorithms that can efficiently leverage modern heterogeneous computational hardware setups that encompass both central processing units (CPUs) and accelerators such as graphical processing units (GPUs).

The main source of computational cost in SCF methods is the repeated construction of the Coulomb and Exchange contributions to the Fock matrix in every iteration.[6, 7] This step involves calculating four-center two-electron repulsion integrals (ERIs) over $N$ Gaussian atomic orbitals (AOs) and subsequently contracting them with the electron density matrix leading to a formal scaling of $O(N^4)$. To avoid storage of the full four index ERI tensor, modern quantum chemistry codes recalculate the ERIs in every iteration and contract them with the density matrix on-the-fly, which presents a significant computational challenge.[8] To make this integral direct approach tractable, extensive research has led to the development of highly efficient integral evaluation[9-11] and screening techniques that achieve $O(N^2)$ or even $O(N)$ empirical scaling when sparsity in the density matrix is also considered.[8, 12-25] Furthermore, the use of robust initial guesses for the density matrix[26, 27] and convergence acceleration techniques,[28-31] including the popular direct inversion in iterative subspace (DIIS), drastically reduce the number of SCF iterations required and thereby the computational cost of the calculation. Coupled with efficient utilization of GPUs in modern codes, SCF methods can nowadays routinely be applied to systems containing thousands of atoms and tens of thousands of basis functions.[22, 23, 32-36]

Nevertheless, integral direct SCF methods suffer from the high prefactor associated with continuous recalculation of ERIs, making such schemes computationally expensive. To avoid using an integral-direct SCF approach, it is necessary to reduce the $O(N^4)$ storage requirements of the ERIs in some way. This has previously been done by decomposing the ERIs into smaller tensors that can be recombined to form the ERIs using a series of elementary operations. The most popular approaches in this category include density fitting,[37, 38] Cholesky decomposition,[39-42] and pseudospectral methods.[43-50] Density fitting and Cholesky decomposition both reduce the storage requirements to $O(N^3)$, but neither provide enough mathematical flexibility to lower the formal computational scaling of the exchange contribution below $O(N^4)$. Pseudospectral methods on the other hand achieve both lowered storage requirements of $O(N^3)$ and a lowered formal computational scaling of $O(N^3)$. However, they do not preserve the permutational symmetry of the integrals and accurate calculations require very large grids (and therefore a substantial prefactor).

Alternatively, one could utilize the more recent tensor hypercontraction (THC) approximation[51-66] that addresses many of the limitations associated with density fitting, Cholesky decomposition, and pseudospectral approaches. Since the THC approximation effectively uncouples all four AO indices and decomposes the ERIs into a product of matrices, the storage requirements are reduced to $O(N^2)$ and the formal scaling of the SCF calculation can be reduced to $O(N^3)$, all while preserving the permutational symmetry of the ERIs. However, the cost of constructing the THC decomposition for molecular systems scales as $O(N^5)$ using conventional ERIs or $O(N^4)$ with density fitting in the standard least-squares fitting on a spatial grid.[53] We note that $O(N^3)$ scaling THC construction has already been realized for periodic system when plane wave basis sets and periodic boundary conditions are used.[66-69] Nevertheless, the $O(N^4)$ scaling of THC construction in the molecular case effectively renders THC unable to lower the cost of SCF calculations for larger molecular systems, as the cost of THC construction will eventually become as expensive as the traditional SCF calculation.

To alleviate this problem, we recently proposed an alternative scheme for constructing the THC decomposition, where we instead fit the central metric matrix to two center electron repulsion integrals in an auxiliary basis set.[62] The resulting THC construction formally only scales as $O(N^3)$ with system size making it practical for SCF methods. Using the new decomposition, we showed that one can devise an accurate THC-SCF method that can potentially provide significant speedups over existing SCF algorithms and thereby enable faster quantum chemical simulations.

In this paper, we devise an efficient GPU-accelerated implementation of the THC-SCF method outlined in our previous letter.[62] We implement THC-SCF in the TeraChem quantum chemistry package[70, 71] to explicitly demonstrate that it provides significant speedup over existing integral direct SCF methods with little sacrifice of accuracy for relative energies. We start by outlining the theory behind the THC-SCF method, followed by a description of our implementation. We then demonstrate that the THC construction is sufficiently fast and accurate to enable a THC based Fock matrix construction that greatly accelerates each SCF iteration and thereby the overall SCF calculation for system sizes up to 5000 basis functions using a single NVIDIA A100 GPU.

**Theory**

In SCF theory,[6, 7] the energy and wavefunction are variationally determined by solving the generalized eigenvalue equation

$$\boldsymbol{FC} = \boldsymbol{SC}\epsilon \tag{1}$$

where $\boldsymbol{F}$ is the AO Fock matrix, $\boldsymbol{C}$ contains the molecular orbital (MO) coefficients used to transform from AO to MO basis, $\boldsymbol{S}$ is the AO overlap matrix, and $\epsilon$ is a diagonal matrix containing the molecular orbital energies. The primary computational expense associated with solving Eq. (1) is the construction of the Fock matrix which involves evaluating a large number of ERIs over $N$ Gaussian AOs $\phi(\vec{r})$:

$$(\mu\nu|\lambda\sigma) = \iiint d\vec{r}_1 \iiint d\vec{r}_2 \, \phi_\mu(\vec{r}_1)\phi_\nu(\vec{r}_1) \frac{1}{r_{12}} \phi_\lambda(\vec{r}_2)\phi_\sigma(\vec{r}_2) \tag{2}$$

leading to a formal scaling of $O(N^4)$. Using the ERIs, the Fock matrix can, in the context of Hartree-Fock theory, be constructed as

$$F_{\mu\nu} = h_{\mu\nu} + J_{\mu\nu} - \frac{1}{2}K_{\mu\nu} \tag{3}$$

where $h_{\mu\nu}$ is the core Hamiltonian, which includes the electron kinetic energy integrals and electron-nuclear attraction integrals, while $J_{\mu\nu}$ and $K_{\mu\nu}$ are the elements of the Coulomb and Exchange matrices, respectively, given as

$$J_{\mu\nu} = \sum_{\lambda\sigma}(\mu\nu|\lambda\sigma)D_{\lambda\sigma} \tag{4}$$

$$K_{\mu\nu} = \sum_{\lambda\sigma}(\mu\lambda|\nu\sigma)D_{\lambda\sigma} \tag{5}$$

with $D_{\lambda\sigma}$ being the density matrix constructed as

$$D_{\mu\nu} = 2 \sum_i C_{i\mu} C_{i\nu} \qquad (6)$$

for a closed-shell system. Since the Fock matrix (more specifically **J** and **K**) depends on the MO coefficients through the density matrix, Eq. (1) must be solved iteratively. This entails either storage or repeated recalculation of the ERIs. Explicit storage of the ERIs is infeasible for larger systems, since the ERI tensor would require approximately 80 TB of memory for a molecule with 1000 basis functions. Modern quantum chemistry codes therefore employ an integral-direct approach,[8] recalculating the ERIs in every iteration and contracting them on-the-fly with the density matrix to build **J** and **K** according to Eqs. (4) and (5), which incurs a substantial computational cost.

One approach that can lower the storage requirements and thus avoid repeated ERI calculation is to decompose the ERIs over components $P$ and $Q$ in the THC format[51-54]

$$(\mu\nu|\lambda\sigma) \approx \sum_P \sum_Q X_\mu^P X_\nu^P Z^{PQ} X_\lambda^Q X_\sigma^Q \qquad (7)$$

thereby reducing the storage requirements to $O(N^2)$. Building on the previous example of 1000 basis functions, the ERIs would now require roughly 1 GB of memory if the number of components (the range of the $P$ and $Q$ indices) is ten times the number of basis functions. Using THC, the ERIs can easily fit into the memory of a single GPU for small to medium sized systems, and the formal scaling of the SCF calculation is reduced from $O(N^4)$ to $O(N^3)$, as unpinning all four AO indices introduces greater mathematical flexibility. A THC-SCF implementation therefore has the potential to provide significant speedup relative to conventional integral-direct SCF schemes and become the new default algorithm for SCF calculations. However, this has not yet occurred, because the state-of-the-art THC construction technique[53] that minimizes the cost function

$$O = \frac{1}{2} \left\| (\mu\nu|\lambda\sigma) - \sum_P \sum_Q X_\mu^P X_\nu^P Z^{PQ} X_\lambda^Q X_\sigma^Q \right\|_F^2 \tag{8}$$

where $F$ denotes the Frobenius norm, using least-squares (LS) fitting on a weighted spatial grid with $X_\mu^P$ given as

$$X_\mu^P = (\omega_P)^{\frac{1}{4}} \phi_\mu(\vec{r}_P) \tag{9}$$

has a formal scaling of $O(N^5)$ using full four center ERIs or $O(N^4)$ using DF integrals. This renders THC unable to speed up SCF calculations, since it has the same (or worse) formal scaling as the SCF calculation itself.

In a recent publication,[62] we showed that the scaling of the THC construction can be reduced to $O(N^3)$ by obtaining an approximation $\tilde{Z}^{PQ}$ to the central metric matrix $Z^{PQ}$ from a fit to two-center Coulomb integrals

$$(A|B) = \iiint d\vec{r}_1 \iiint d\vec{r}_2 \, \chi_A(\vec{r}_1) \frac{1}{r_{12}} \chi_B(\vec{r}_2) \tag{10}$$

in an auxiliary basis $\chi_A(\vec{r})$. In the case where $\chi_A$ approximately covers the function space of the pair products of regular basis functions, i.e. $\phi_\mu \phi_\nu$ and $\phi_\lambda \phi_\sigma$, this procedure will give $\tilde{Z}^{PQ}$ as a good approximation to $Z^{PQ}$. Fundamentally, this is the same approximation used in density fitting, which also relies on highly optimized auxiliary basis sets. However, in this case we face the slight disadvantage that the method does not directly minimize the error in the ERIs within a given auxiliary basis set. Instead, the error in $(A|B)$ is minimized meaning that the importance of auxiliary basis set is amplified. Following this approach, $\tilde{Z}^{PQ}$ can be determined by computing the THC decomposition of $(A|B)$ as

$$(A|B) \approx \sum_P \sum_Q X_A^P \tilde{Z}^{PQ} X_B^Q \tag{11}$$

where $X_A^P$ are collocation matrices defined on the grid as

$$X_A^P = (\omega_P)^{\frac{1}{2}} \chi_A(\vec{r}_P) \tag{12}$$

Following the LS-THC procedure,[53] this can be done by minimizing the cost function

$$O = \frac{1}{2} \left\| (A|B) - \sum_P \sum_Q X_A^P \tilde{Z}^{PQ} X_B^Q \right\|_F^2 \tag{13}$$

with respect to $\tilde{Z}^{PQ}$ in a least-squares sense giving

$$Z^{PQ} \approx \tilde{Z}^{PQ} = \sum_{P'} \sum_{Q'} \left( \sum_A X_A^P X_A^{P'} \right)^{-1} X_A^{P'} (A|B) X_B^{Q'} \left( \sum_B X_B^Q X_B^{Q'} \right)^{-1} \tag{14}$$

where all involved matrices are of size $O(N^2)$ meaning that the necessary matrix multiplications and matrix (pseudo-)inverse only require $O(N^3)$ floating point operations. In Eq. (14), the grid metric matrix defined by $\left( \sum_A X_A^P X_A^{P'} \right)$ is nearly singular, as the number of grid points required is generally 2-3 times larger than the number of auxiliary basis functions, necessitating a pseudoinversion. This has been the limiting step in previous LS-THC implementations.[52, 59] However, the pseudoinversion can be circumvented by considering a simpler cost function

$$O = \frac{1}{2} \left\| L_{AC} - \sum_P X_A^P Y_C^P \right\|_F^2 \tag{15}$$

where $L_{AC}$ can e.g. be obtained from a complete Cholesky decomposition of $(A|B)$

$$(A|B) = \sum_C L_{AC} L_{BC} \tag{16}$$

Minimizing Eq. (15) with respect to $Y_C^P$ gives

$$\sum_P X_A^P Y_C^P = L_{AC} \tag{17}$$

which represents sets of underdetermined linear systems from which we can determine a solution $Y_C^P$ and obtain $\tilde{Z}^{PQ}$ and thereby an approximation to $Z^{PQ}$ as

$$Z^{PQ} \approx \tilde{Z}^{PQ} = \sum_C Y_C^P Y_C^Q \tag{18}$$

The reformulated cost function in Eq. (15) achieves two key advantages. Not only is the coefficient matrix $X_A^P$ much smaller in size than $\left(\sum_A X_A^P X_A^{P'}\right)$ but crucially we also avoid the pseudoinversion of $\left(\sum_A X_A^P X_A^{P'}\right)$. As a result, direct solution of Eq. (17) with standard techniques (e.g., LQ decomposition for underdetermined systems) is both more numerically stable and more computationally efficient compared to Eq. (14).[72]

**Implementation Details**

We have developed a GPU-accelerated implementation of the methodology described above in the TeraChem quantum chemistry package.[70, 71] The current implementation is optimized to run on a single node using a single GPU which provides excellent performance for systems containing up to ~5000 basis functions, where the $X_\mu^P$ and $Z^{PQ}$ tensors can be stored in GPU memory of newer GPUs using around 20 GB of memory in double precision. As outlined in our previous paper, we combine the THC approximation with the previously reported density difference SCF scheme[73] to improve the accuracy. Given that we have not optimized basis sets or grids specifically for this new THC decomposition scheme, the approximation made in Eq. (18) remains the limiting factor in terms of accuracy and it surely leaves room for improvement. We minimize the effect of THC errors by employing a SAD initial guess and a few subsequent conventional iterations in double precision until the energy change is below 0.1 $E_h$, at which point the density is taken as the reference density for the remainder of the SCF iterations using THC integrals. The current implementation does not exploit integral screening or sparsity in the density matrix and we thus compute and use every element of the $X_\mu^P$, $X_A^P$ and $Z^{PQ}$ tensors in double precision. Future work will explore the possibility of improving the performance by taking advantage of sparsity and mixed precision.[32]

In **Algorithm 1**, we outline our implementation of the THC construction with $O(N^3)$ formal scaling. Initially, $X_A^P$ and $(A|B)$ are computed in steps 1 and 2 using custom CUDA kernels and hand optimized quadrature grids developed in previous papers on LS-THC that are available in TeraChem.[56, 74] Following this, $X_A^P$ and $(A|B)$ are transferred onto the GPU where we first perform a complete Cholesky decomposition of $(A|B)$ in step 3 to form the right-hand side of Eq. (17). Subsequently, the underdetermined sets of linear equations in Eq. (17) are solved on the GPU using an LQ decomposition of $X_A^P$ to obtain $Y_C^P$ and ultimately $Z^{PQ}$. Lastly, we compute $X_\mu^P$ using the same kernels as in step 1 and for the remainder of the THC-SCF calculation $X_\mu^P$ and $Z^{PQ}$ are stored in GPU memory. The implementation of **Algorithm 1** utilizes NVIDIA's cuSOLVER and cuBLAS libraries for all linear algebra computations involved in steps 3-5 to leverage the highly optimized routines provided therein.

| | |
|---|---|
| 1: | Form $X_A^P = (\omega_P)^{\frac{1}{2}} \chi_A(\vec{r}_P)$ |
| 2: | Form $(A\|B)$ |
| 3: | Compute Cholesky decomposition: $(A\|B) = \sum_C L_A^C L_B^C$ |
| 4: | Solve $\sum_P X_A^P Y_C^P = L_A^C$ |
| 5: | Form $Z^{PQ} = \sum_C Y_C^P Y_C^Q$ |
| 6: | Form $X_\mu^P = (\omega_P)^{\frac{1}{4}} \phi_\mu(\vec{r}_P)$ |

**Algorithm 1.** Pseudocode for the $O(N^3)$ scaling THC construction.

In **Algorithm 2** and **Algorithm 3**, we outline our implementation of the THC-J-build and the THC-K-build, respectively. For both the THC-J-build and THC-K-build, we initially transfer the current density matrix for which the $J$ or $K$ matrix is to be constructed from CPU memory to GPU memory. Following this, steps 2-6 in both algorithms perform a series of tensor contractions to build either $J$ or $K$ with $O(N^3)$ formal scaling. In all steps where it is possible, **Algorithm 2** and **Algorithm 3** utilize cuBLAS for matrix-matrix multiplications. However, some of the necessary steps are not standard matrix-matrix multiplications and therefore require custom CUDA kernels. More specifically, custom CUDA kernels are used for the nested dot product in step 3 of **Algorithm 2**, the nested Hadamard product in step 5 of **Algorithm 2**, and the Hadamard product in step 4 of **Algorithm 3**. Following the formation of $J$ and/or $K$, the remaining steps of the SCF calculation such as Fock matrix diagonalization and DIIS interpolation are carried out using the existing infrastructure for standard SCF calculations in TeraChem, as these already formally scale as $O(N^3)$ with system size.

| | |
|---|---|
| 1: | Transfer $D_{\lambda\sigma}$ from CPU to GPU |
| 2: | Form $A_\lambda^Q = \sum_\sigma D_{\lambda\sigma} X_\sigma^Q$ |
| 3: | Form $B^Q = \sum_\lambda A_\lambda^Q X_\lambda^Q$ |
| 4: | Form $C^P = \sum_Q Z^{PQ} B^Q$ |
| 5: | Form $E_\nu^P = X_\nu^P C^P$ |
| 6: | Form $J_{\mu\nu} = \sum_P X_\mu^P E_\nu^P$ |
| 7: | Transfer $J_{\mu\nu}$ from GPU to CPU |

**Algorithm 2.** Pseudocode for the J-build using THC assuming $X_\mu^P$ and $Z^{PQ}$ are already allocated on the GPU.

| | |
|---|---|
| 1: | Transfer $D_{\lambda\sigma}$ from CPU to GPU |
| 2: | Form $A_\sigma^Q = \sum_\lambda X_\lambda^Q D_{\lambda\sigma}$ |
| 3: | Form $B^{PQ} = \sum_\sigma X_\sigma^P A_\sigma^Q$ |
| 4: | Form $C^{PQ} = Z^{PQ} B^{PQ}$ |
| 5: | Form $E_\nu^P = \sum_Q C^{PQ} X_\nu^Q$ |
| 6: | Form $K_{\mu\nu} = -\frac{1}{2}\sum_P X_\mu^P E_\nu^P$ |
| 7: | Transfer $K_{\mu\nu}$ from GPU to CPU |

**Algorithm 3.** Pseudocode for the K-build using THC assuming $X_\mu^P$ and $Z^{PQ}$ are already allocated on the GPU.

**Computational Details**

In the following, we show Hartree-Fock results obtained using the GPU-accelerated THC-SCF implementation in TeraChem that was outlined in the previous section, along with comparisons to the existing standard SCF code in TeraChem. All calculations utilize the cc-pVDZ basis set,[75] the corresponding cc-pVDZ-RI auxiliary basis set,[76] and the R-DVR grids that were developed and optimized in previous studies.[56, 74] As alluded to earlier, we refine the initial density using full precision iterations prior to starting the THC-SCF calculation in the density difference scheme. Below, we will explicitly state the number of iterations used in each case. All timings presented below were obtained from calculations performed on a single NVIDIA DGX-A100 node using a single NVIDIA A100 GPU and two 64-core 2.25 GHz AMD EPYC 7742 CPUs.

**Results and Discussion**

To test the performance of the developed THC-SCF implementation, we perform a series of calculations on $(H_2O)_n$ clusters with n=10-100 (250-2500 basis functions) using a single standard SCF iteration for initial refinement. Our aim is to illustrate the cost of computing the THC factorization using **Algorithm 1** and the extent to which the THC factorization speeds up the J- and K-builds relative to the standard implementation in TeraChem. In **Figure 1**, we report a wall time breakdown of the developed THC construction. As expected, the most expensive part of the THC construction is the formation of $Z^{PQ}$ according to steps 3-5 of **Algorithm 1**, which despite the empirically observed $O(N^2)$ scaling, formally scales as $O(N^3)$. However, the cost of constructing $X_A^P$ in the auxiliary basis set is also seen to take up a significant amount of time, since the auxiliary basis set contains roughly five times as many basis functions as the primary basis set. Meanwhile, the formation of $X_\mu^P$ and $(A|B)$ is relatively fast in this context and we do not expect these steps to become a bottleneck for any system size. Overall, we observe an empirical $O(N^2)$ scaling for the THC formation, which is significantly lower than the expected $O(N^3)$ scaling. We attribute this lower empirical scaling to two factors: 1) A system size of 2500 basis functions is likely not large enough for the linear system in step 4 of **Algorithm 1** to reach the asymptotic scaling regime and 2) a non-negligible portion of the wall time is spent computing $X_A^P$, $X_\mu^P$, and $(A|B)$ that actually have a formal scaling of $O(N^2)$. Using our current implementation, the THC factorization can be computed in less than 8 seconds for the $(H_2O)_{100}$ cluster which corresponds to the cost of less than 2 conventional SCF iterations (see **Figure 2**) and thus represents a significant speedup relative to previous least-squares THC implementations with $O(N^4)$ scaling.[53, 59, 60]

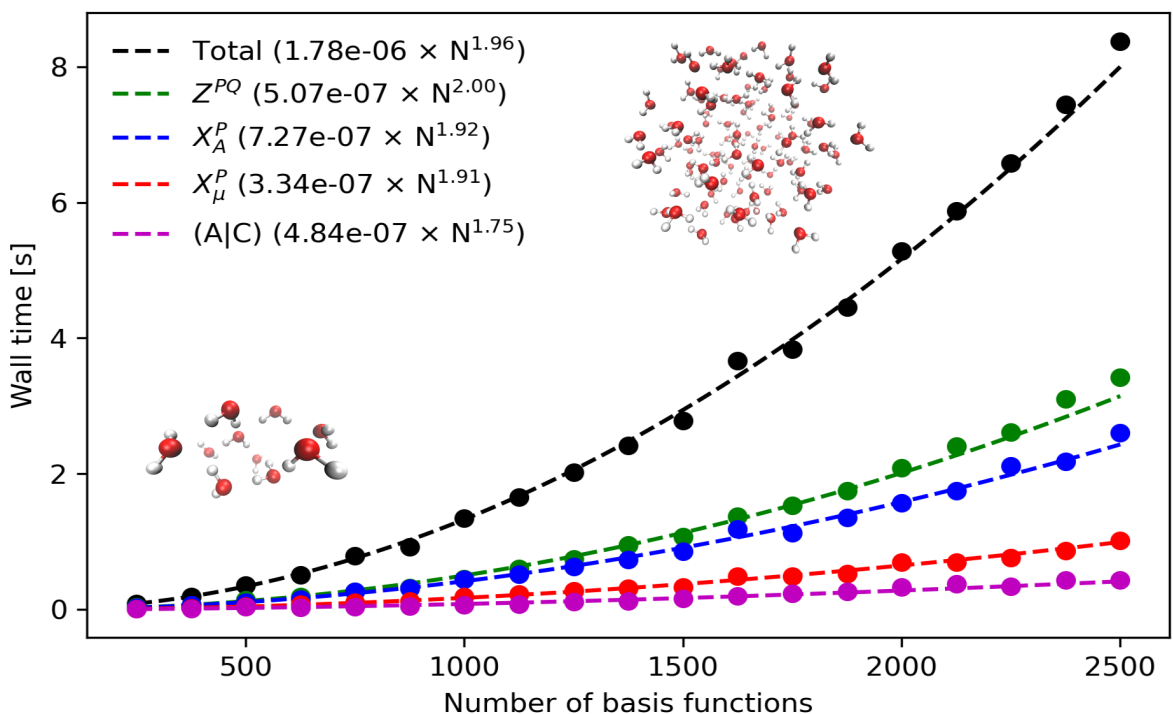

**Figure 1.** Timing breakdown of the THC construction outlined in **Algorithm 1** for a series of water clusters ranging from 10 to 100 water molecules. The shown fits are from a linear fit on a double logarithmic scale. Note that $Z^{PQ}$ formation corresponds to steps 3-5 in **Algorithm 1**.

Having developed a sufficiently fast THC construction, we now turn to comparing the wall time obtained for the J- and K-builds using conventional integrals and using THC as outlined in **Algorithm 2** and **Algorithm 3**. In **Figure 2**, we display the average wall time for an SCF iteration, a J-build, and a K-build for the standard SCF scheme and the THC-SCF scheme in TeraChem. It should be noted that the THC-SCF iteration times do not account for the construction of the reference density used in the density difference SCF scheme, which corresponds to a full precision standard SCF iteration. As seen from **Figure 2**, the THC-SCF iterations show a similar empirical scaling compared to standard SCF and they are in general much faster than the corresponding standard SCF iterations, achieving more than a twofold speedup for the larger systems owing to the lower prefactor obtained by circumventing repeated ERI calculation. As seen in previous studies,[64, 65] the main performance increase comes from the two- to fourfold speedup achieved in the K-build, for which THC enables a lower $O(N^3)$ formal scaling and a very straightforward pattern of linear algebra operations as seen in **Algorithm 3**, resulting in a prefactor that is one order of magnitude lower than the standard K-build.

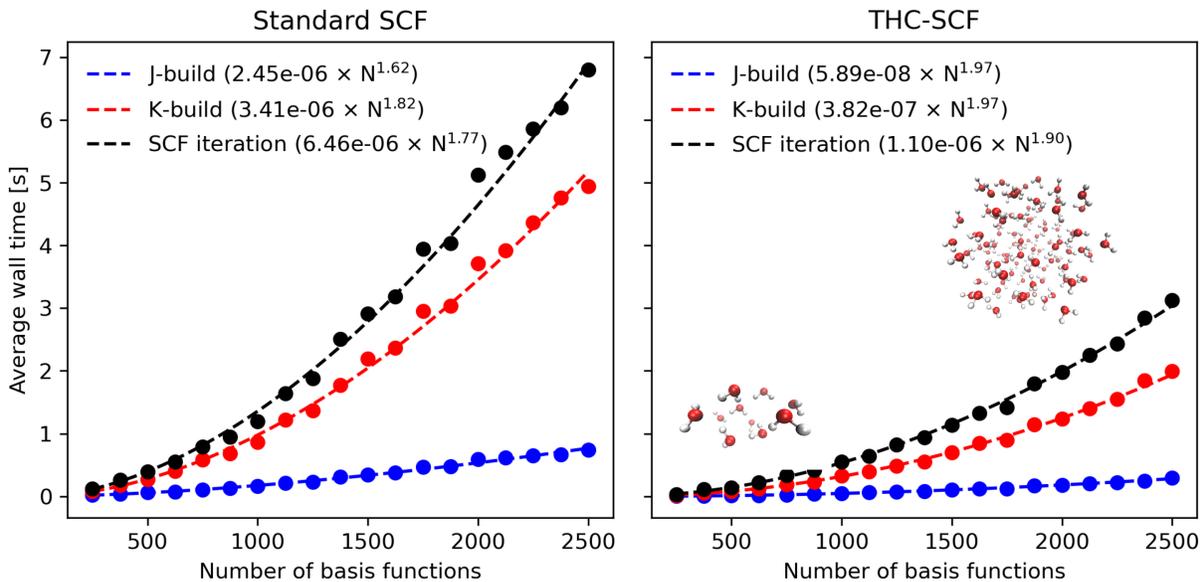

**Figure 2.** Average wall time for an SCF iteration, a J-build and a K-build for a series of water clusters ranging from 10 to 100 water molecules. (Left) Standard SCF timings and (Right) THC-SCF timings excluding the full precision refinement iteration in THC-SCF. The shown fits are from a linear fit on a double logarithmic scale.

For the J-build, we also observe a similar two- to fourfold speedup using THC despite a higher empirical scaling of $O(N^2)$ compared to the $O(N^{1.6})$ scaling of the standard SCF scheme. This can again be attributed to the significantly lower prefactor. However, the effects of speeding up the J-build are much more limited, since the standard J-build is already highly optimized and does not represent the main bottleneck in standard Hartree-Fock and hybrid DFT calculations. Overall, it is evident that THC can provide significant acceleration of SCF iterations by reducing the wall time of average iterations by more than 50%. Interestingly, this is achieved using conceptually very simple THC-J- and THC-K-build implementations that do not exploit integral screening, density matrix sparsity, or dynamic precision. Exploiting any or all of these could provide even greater performance. In contrast, the state-of-the-art GPU accelerated J- and K-build implementations relies on heavily optimized hand-crafted GPU kernels that require a tremendous effort to optimize.[22, 23, 33]

Having shown that the THC-SCF scheme can provide significant speedup over existing SCF implementations, it remains to test whether the THC-SCF scheme can reliably produce electronic energies that are accurate enough to model relative energy profiles. To this end, we initially study the relative energy landscape of alanine dipeptide obtained from scanning over the $\psi$ and $\phi$ dihedral angles shown in **Figure 3c** with 24 steps in each, yielding a total of 576 structures. Figures 3a-b display the resulting relative energy landscape obtained using standard SCF and THC-SCF. Figure 3d shows the errors of the THC-SCF method obtained using a single conventional iteration for initial refinement. It is evident from **Figure 3** that THC-SCF provides a relative energy landscape for alanine dipeptide that is visually almost indistinguishable from that obtained using standard SCF. Looking at the errors, we note that all relative energies are accurate within 1.1 kcal/mol and the root mean square error is 0.26 kcal/mol, meaning that the THC-SCF method captures the underlying physics in its current form. While the errors are slightly higher than those regularly seen in e.g. density fitting,[38, 76] it should still be reiterated that we currently have not optimized auxiliary basis sets for the approximation made in Eq. (18), and that unlike in density fitting, this THC scheme does not provide a fit that directly minimizes the error of the obtained ERIs within the given auxiliary basis set. We therefore expect that the accuracy of THC-SCF can be improved in future studies by rigorously tailoring the auxiliary basis set to the approximation in Eq. (18).

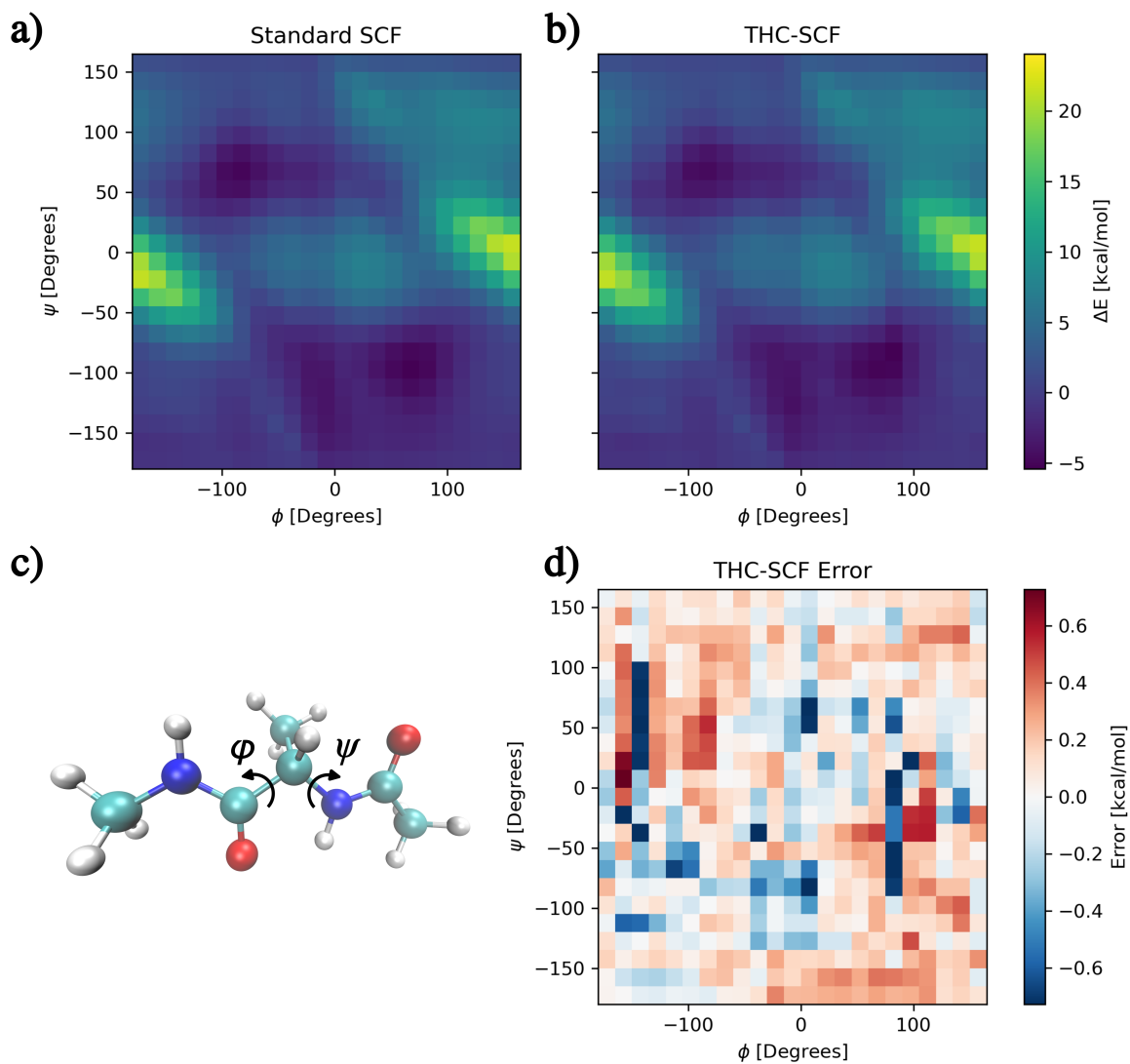

**Figure 3.** Relative energy landscape of alanine dipeptide obtained from scanning over the $\psi$ and $\phi$ dihedral angles shown in panel c) in 24x24 steps yielding 576 geometries. Panels a) and b) show the results of standard SCF and THC-SCF using one standard iteration for initial guess refinement, respectively. Panel d) shows the errors of the THC-SCF scheme relative to the standard SCF results.

Finally, we apply the THC-SCF methodology to model the relative energy landscape along a minimum energy path connecting the $3_{10}$-helical and $\alpha$-helical structures of the HIV-1 epitope (PDB ID: 1LB0) following previous work.[77] We obtain a minimum energy path using the nudged elastic band method[78] with eight frames obtained from a geodesic interpolation[79] connecting the two optimized endpoints. Here, our objective is to compare the results to those of standard SCF to validate the potential of using the current THC-SCF scheme for modeling the energetics of smaller proteins. **Figure 4** depicts the energy profiles obtained using standard SCF and THC-SCF as well as the errors of the THC-SCF scheme For these calculations, four refinement iterations were used, as the total energy for systems like HIV-epitope is of the order of $10^4$ hartree demanding a highly precise initial guess to reduce energy changes below 0.1 $E_h$ to reduce numerical accuracy requirements of the THC factorization.

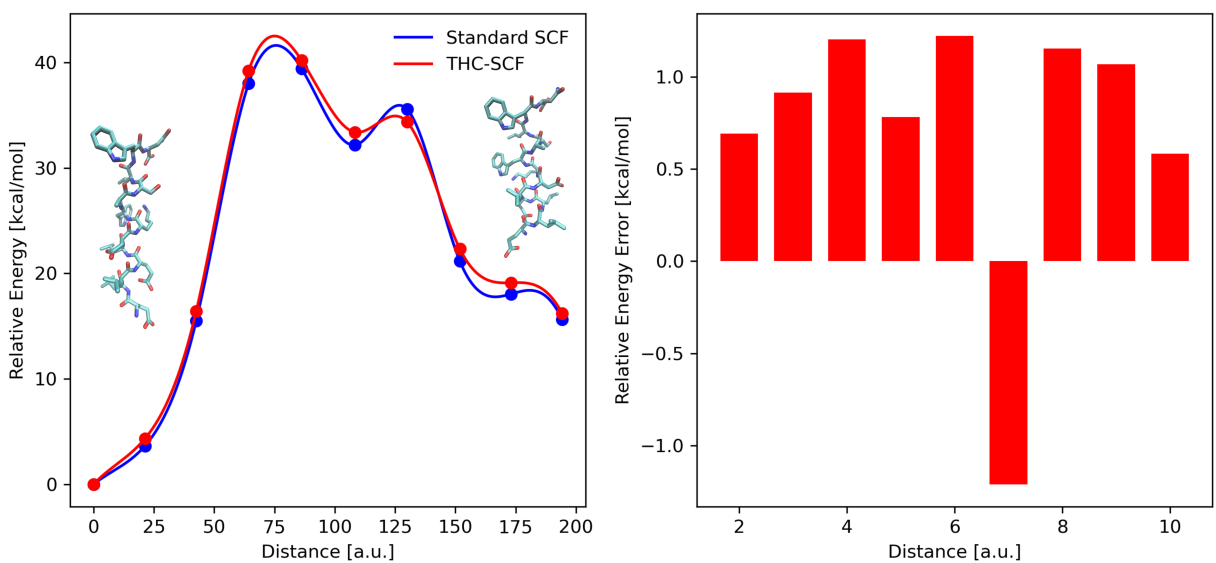

**Figure 4.** (Left) Relative energy profile connecting the $3_{10}$-helical and $\alpha$-helical structures of the HIV-1 epitope obtained using standard SCF and THC-SCF. (Right) Errors in the relative energies for the THC-SCF scheme obtained using four standard iterations for initial guess refinement.

**Figure 4** highlights that THC-SCF qualitatively captures the relative energy profile of the HIV-1 epitope with minor visual differences, which was also the case for alanine dipeptide in **Figure 3**. The errors in relative energies show that the current implementation of THC-SCF has an RMSE of 0.96 kcal/mol with errors ranging from -1.21 kcal/mol to 1.22 kcal/mol. Again, we note that the errors in the relative energies are slightly larger than those usually seen in e.g. density fitting, highlighting the need for tailored auxiliary basis sets to improve the applicability of the THC-SCF scheme.

**Conclusion**

In this paper, we devised a GPU-accelerated implementation of the formally $O(N^3)$ scaling THC-SCF methodology reported in our recent letter.[62] Using our new approach for constructing the THC factorization by fitting to two-center Coulomb integrals in an auxiliary basis as a proxy for the standard LS-THC fit to ERIs, the THC fit can be determined with $O(N^2)$ empirical scaling at a cost corresponding to less than two standard SCF iterations. This feat now enables the usage of THC in the context of SCF methods, since THC provides significant speedup of both the J- and K-builds and thereby SCF iterations by a factor of 2-4 for systems containing thousands of basis functions compared to the standard SCF algorithm in TeraChem. Not only is the THC-SCF scheme faster than the standard SCF scheme, but it is also conceptually much simpler to implement on a GPU, as most performance-critical operations are relatively simple linear algebra operations available through optimized libraries. Adding to this point, it is noticeable that the observed speedup is achieved without exploiting integral screening, sparsity in the density matrix or dynamic precision in the THC-J- and THC-K-builds. It is solely a product of the much lower memory footprint of the THC factor matrices that allows us to circumvent the repeated recalculation of ERIs leading to a much lower computational prefactor. In the future, further speedup of the THC-SCF scheme can surely be obtained by utilizing integral screening, sparsity in the density matrix as well as dynamic precision, which has successfully led to major algorithmic advances in standard integral direct SCF schemes.

By utilizing the density difference SCF scheme, the developed THC-SCF method enables modelling of energy differences for small to medium sized protein systems with kcal/mol accuracy. As shown by our sample calculations, this accuracy allows us to obtain qualitatively accurate potential energy surfaces. However, in its current form that reuses auxiliary basis sets optimized for density fitting, we find that reaching quantitative precision of relative energies for larger systems such as TrpCage and beyond requires further development of the new THC construction. Our observations leave the auxiliary basis set used for the THC construction as the fundamental limitation for the accuracy of the THC factorization and thus the method. To accurately reproduce the ERIs, it is essential that the auxiliary basis set covers the function space of the pair product space of the primary basis functions. In future studies, we will devise strategies for improving the accuracy of the THC factorization used for THC-SCF to enable its use as a promising full replacement for conventional ERIs.

**Supporting Information**

The geometries used for all test cases as well as raw energies and timings used to produce all figures presented in the paper.

**Notes**

TJM is a founder of PetaChem, LLC.

**Acknowledgements**

This work was supported by the Office of Naval Research (N00014-21-1-2151). AEHB acknowledges financial support from the Novo Nordisk Foundation under grant reference number NNF24OC0089345.